\documentstyle[times,floats,aps,prb]{revtex}

\newcommand{\be}{\begin{eqnarray}}
\newcommand{\ee}{\end{eqnarray}}

\newcommand{\bmath}{\begin{mathletters}}
\newcommand{\emath}{\end{mathletters}}

\begin{document}
\input{epsf}

\title {\bf Optimization of quantum Monte Carlo wave functions using analytical energy derivatives}
\author{Xi Lin, Hongkai Zhang and Andrew M. Rappe}
\address{Department of Chemistry and Laboratory for Research on the Structure of Matter, University of Pennsylvania, Philadelphia, PA 19104}
\date{\today}
\maketitle

\begin{abstract} 

An algorithm is proposed to optimize quantum Monte Carlo (QMC) wave functions based on Newton's method and analytical computation of the first and second derivatives of the variational energy. This direct application of the variational principle yields significantly lower energy than variance minimization methods when applied to the same trial wave function. Quadratic convergence to the local minimum of the variational parameters is achieved. A general theorem is presented, which substantially simplifies the analytic expressions of derivatives in the case of wave function optimization. To demonstrate the method, the ground state energies of the first-row elements are calculated.

\end{abstract} 

\section{\bf INTRODUCTION}

Quantum Monte Carlo is a powerful method of solving the Schr\"{o}dinger equation. QMC treats many-body correlation in an efficient and flexible way, enabling highly accurate studies of atoms, small molecules and clusters. \cite{{njr:lester},{njr:raghavachari},{njr:ceperley}} A high-quality trial wave function is crucial to the calculation, since the trial function determines the ultimate accuracy one can achieve in variational Monte Carlo (VMC) and fixed-node diffusion Monte Carlo, and trial function quality dramatically affects the efficiency of the computation.

An algorithm which efficiently and reliably optimizes wave functions is a critical tool for VMC calculations. One straightforward approach for improving the VMC wave function is to perform energy minimization, in which the variational parameters are altered with the goal of lowering the expectation value of the energy. This approach is complicated in VMC because of the uncertainties associated with stochastic sampling. In order to determine whether a new set of parameters yields a lower energy than the current set, one needs to sample a large number of configurations to ensure that the energy difference between the two sets of parameters is actually larger than the energy error bars. Correlated sampling methods are frequently performed to improve the efficiency of energy minimization. Typically, the energy is calculated using identical sampling points in configuration space for two trial wave functions which differ by a single parameter. The process is repeated for each parameter, and steepest-descent techniques are commonly used for parameter updating. \cite{njr:shuang} This correlated sampling approach requires a significant amount of memory (to store data for every sampling point) and the numerical differentiation $\Delta E / \Delta c$ requires many extra evaluations of the local energy. For systems with a large number of parameters, numerical evaluation of the required derivatives becomes computationally intractable. Analytical energy derivative techniques are very seldom used in current VMC calculations. We will concentrate on this in the following sections.

A successful alternative approach has been developed which focuses on lowering the variance of the local energy, $\widehat{H}\Psi/\Psi$. \cite{njr:bartlett} If the wave function $\Psi$ were the exact ground eigenstate, the local energy would be a constant with a variance of zero. A major strength of the variance minimization approach is that the quantity to be minimized has a minimum value which is known {\it a priori} (unlike energy minimization). This idea has been implemented in various ways and has recently become a nearly universal approach in VMC wave function optimizations. Typically, one calculates first derivatives of the local energy variance analytically. Steepest-descent techniques \cite{{njr:hhuang},{njr:hhuang2}} or a combination of analytic first derivatives with approximate expressions for the second derivatives are then used for wave function variance reduction (a least-squares fit). \cite{{njr:buckert},{njr:schmidt},{njr:luchow}} Although variance methods have the remarkable strength of an {\it a priori} minimum value of zero, it is much harder to compute the first and second derivatives of the variance analytically compared to variational energy methods. Therefore, approximate analytical derivatives beyond the first-order are used in real calculations, and to our knowledge the validity of these approximations has not been discussed within the scope of VMC wave function optimization. It is important to point out that the ``direction sets'' minimum-searching methods, such as steepest-descent and conjugate gradient are not efficient for wave function optimization in VMC, because these line-minimization techniques require at least one order of magnitude more evaluations of local energy along the search directions. Moreover, variance minimization is actually an indirect method, since a smaller variance does not necessarily correspond to a lower energy, and the main goal of variational methods such as VMC is the lowest possible upper bound to the energy.

Correlated sampling can be used (instead of analytic derivatives) to lower the variance of the local energy. One excellent version of this idea is known as the fixed-sample method. \cite{njr:umrigar} In this approach, the sampling points for the objective function (variance of the local energy in this case) are fixed during the optimization procedure, which makes it possible to reduce stochastic noise during the optimization. In addition, it has been observed from a few preliminary calculations that the number of configurations sufficient for parameter updating does not increase rapidly with system size. \cite{njr:umrigar} The use of very complex trial correlation functions has yielded highly accurate energies for a few first-row atoms. \cite{{njr:umrigar},{njr:umrigar2}} However, this fixed-sample procedure can have problems if the variational parameters affect the nodes, since the density ratio of the current and initial trial wave functions diverges frequently in the area around the nodes of the trial wave function. Even worse, this density ratio increases exponentially with the size of the system. \cite{njr:kent} Although manually setting an upper bound for the weights or introducing a nodeless sampling probability density function can overcome the singularities in the fixed distribution \cite{njr:barnett}, a general and appropriate form for the positive definite functions is still unavailable. In addition, the variational energy from fixed-sample calculations can be sensitive to the choice of reference energy, sample size, and convergence criteria. \cite{njr:sun} 

The method we present involves updating the variational parameters to lower the energy expectation value, guided by the force vectors and Hessian matrix of the variational energy with respect to variational parameters. Generally it converges quadratically, making it more efficient than the steepest-descent or quasi-Newton techniques employed in the variance minimization procedure. \cite{{njr:hhuang},{njr:buckert}} In most cases, the best set of parameters can be obtained after only one or two iterations. Beginning with an identical trial wave function and the same variational parameters, the correlation energies obtained from our method are significantly better than results in the literature. \cite{njr:schmidt} With this approach, we also demonstrate the ability to optimize a wave function with a large number of parameters. All of the data are collected and compared in Section IV.

\section{\bf VMC AND OPTIMIZATION ALGORITHM} 

Variational Monte Carlo allows us to take our physical insights and construct a trial wave function $\Psi_{\rm T}$ containing a set of variational parameters $\left\{c_{m}\right\}$. The parameters are varied with the goal of reducing the energy expectation value. In VMC, the true ground state energy is given by the Raleigh-Ritz quotient: 
\begin{eqnarray}
E_{0} \leq E_{\rm T}\left(\left\{c_{m}\right\}\right) &=& \frac{\int\Psi_{\rm T}^{*}\left(\left\{c_{m}\right\}\right)\widehat{H}\Psi_{\rm T}\left(\left\{c_{m}\right\}\right) d\tau}{\int\Psi_{\rm T}^{*}\left(\left\{c_{m}\right\}\right)\Psi_{\rm T}\left(\left\{c_{m}\right\}\right) d\tau} \nonumber \\
&=& \lim_{N\rightarrow\infty}\frac{1}{N}\sum_{\alpha = 1 }^{N}\left(E_{\rm L}\right)_{\alpha}, \nonumber
\end{eqnarray}
\noindent where $E_{\rm L} \equiv \widehat{H}\Psi_{\rm T}/\Psi_{\rm T}$ is called the local energy and $\alpha$ is a configuration-space point, visited with relative probability $\Psi^{*}_{\rm T}\Psi_{\rm T}$, the density of the trial wave function at $\alpha$. 

In a bound molecular system with fixed nuclei, the non-relativistic Hamiltonian
$$\widehat{H} = -\frac{1}{2}\sum_{i}\nabla^{2}_{i} - \sum_{i,I}\frac{Z_{I}}{r_{iI}} + \sum_{i<j}\frac{1}{r_{ij}}. $$
\noindent has inversion symmetry. (Note that capital letter subscripts refer to nuclei and lower--case letters refer to electrons.) Therefore, the true ground-state wave function of this class of Hamiltonian can generally be constructed without an imaginary part, i.e., 
$$\Psi^*_{\rm T}\left(\left\{c_{m}\right\}\right)=\Psi_{\rm T}\left(\left\{c_{m}\right\}\right).$$ 

In this case, the expectation value of the energy and the first derivative of energy with respect to a variational parameter can be written as
\begin{eqnarray}
E&=&\frac{\int\Psi\widehat{H}\Psi d\tau}{\int\Psi^{2}d\tau}, \nonumber \\
\frac{\partial E}{\partial c_m}&=&\frac{1}{\int\Psi^{2}d\tau}\left(\int\frac{\partial \Psi}{\partial c_m}\widehat{H}\Psi d\tau+\int\Psi\widehat{H}\frac{\partial \Psi}{\partial c_m}d\tau\right) \nonumber \\
 & &-\frac{1}{\left(\int\Psi^{2}d\tau\right)^2}\int\Psi\widehat{H}\Psi d\tau\int 2\Psi\frac{\partial \Psi}{\partial c_m}d\tau.
\end{eqnarray}
Because 
\begin{eqnarray}
\int\frac{\partial\Psi}{\partial c_m}\widehat{H}\Psi d\tau =\int\Psi\widehat{H}\frac{\partial \Psi}{\partial c_m} d\tau \nonumber, 
\end{eqnarray}
\noindent for real wave functions, we simplify Eq. (1) and obtain
\begin{eqnarray}
\frac{\partial E}{\partial c_{m}}&=&\frac{2}{\int\Psi^{2}d\tau}\int\Psi^2\left(\frac{\widehat{H}\Psi}{\Psi}\right)\left(\frac{\frac{\partial \Psi}{\partial c_{m}}}{\Psi}\right) d\tau \nonumber \\
 & &-\frac{2}{\left(\int\Psi^{2}d\tau\right)^{2}} \int\Psi^{2}\frac{\widehat{H}\Psi}{\Psi}d\tau \int{\Psi}^2\frac{\frac{\partial \Psi}{\partial c_{m}}}{\Psi}d\tau \nonumber \\ 
 &=& \lim_{N\rightarrow\infty}\frac{2}{N}\sum_{\alpha=1}^{N}\left\{\left( E_{\rm L} \times \Psi_{\ln,m}^{'} \right)_\alpha - E \times \left(\Psi_{\ln,m}^{'}\right)_\alpha \right\}, 
\end{eqnarray} 
where we define $$\Psi_{\ln,m}^{'} \equiv \frac{\partial \ln\Psi}{\partial c_{m}} = \frac{\frac{\partial \Psi}{\partial c_{m}}}{\Psi}$$ 
We notice that the finite sum for different terms performed in the same configuration samplings in the formula above can make more efficient computation and reduce the fluctuations in the sense of correlated sampling. 

Similarly, one can compute the second derivatives of variational energy with respect to variational parameters as  
\begin{eqnarray}
& &\frac{\partial^{2}E}{\partial c_{m}\partial c_{n}} \nonumber \\
= & &\lim_{N\rightarrow\infty}\frac{2}{N}\sum_{\alpha=1}^{N}\left\{\left(E_{\rm L} \times \Psi_{\ln,m,n}^{''}\right)_{\alpha}-E\times\left(\Psi_{\ln,m,n}^{''}\right)_{\alpha} \right.\nonumber\\
& & + 2\left[\left(E_{\rm L} \times \Psi_{\ln ,m}^{'}\times \Psi_{\ln ,n}^{'}\right)_{\alpha} -E \times \left(\Psi_{\ln ,m}^{'}\times \Psi_{\ln ,n}^{'}\right)_\alpha \right]\nonumber \\
& &-\left(\Psi_{\ln ,m}^{'}\times\frac{\partial E}{\partial c_{n}}\right)_{\alpha}-\left(\Psi_{\ln ,n}^{'}\times\frac{\partial E}{\partial c_{m}}\right)_{\alpha} \nonumber \\
& &\left.+\left(\Psi_{\ln ,m}^{'} \times E_{{\rm L}, n}^{'} \right)_{\alpha}\nonumber \right\} , \nonumber
\end{eqnarray}

where 
$$\Psi_{\ln,m,n}^{''}= \frac{\partial^2 \ln \Psi}{\partial c_{m} \partial c_{n}}, $$ and
$$E_{{\rm L}, n}^{'}= \frac{\partial E_{\rm L}}{\partial c_{n}} $$ 

We perform a standard Metropolis walk with importance sampling for E and its first and second derivatives. This gives numerical values for the force vector ${\bf b}$ and Hessian matrix ${\bf H}$, which are defined as 
$${\bf b} = \left( \frac{\partial E}{\partial c_{m}} \right) $$ 
and 
$$ {\bf H} = \left(\frac{\partial^{2} E}{\partial c_{m} \partial c_{n} } \right).$$
The parameters are then updated according to
$$ {\bf c}_{\rm next} = {\bf c}_{\rm cur} - {\bf H}^{-1} \cdot {\bf b} $$
\vspace{0.1in}\noindent until converged. Here ${\bf c}_{\rm cur}$ and ${\bf c}_{\rm next}$ stand for the current and next values of the trial parameter set respectively.

\section{\bf THEOREM OF LOCAL OBSERVABLE QUANTITY DERIVATIVE} 

We now demonstrate that the expectation value of the first derivative of the local value ${\cal O}_{\rm L} \equiv \widehat{\cal O}\Psi/ \Psi$ of any Hermitian operator $\widehat{\cal O}$ with respect to any real parameter $c$ in any real wave function $\Psi$ is always zero, i.e.,
\be
\lim_{N\rightarrow\infty}\frac{1}{N}\sum_{\alpha=1}^{N}\left(\frac{\partial {\cal O}_{\rm L}}{\partial c}\right)_\alpha\equiv 0. 
\ee
Explicitly, the left hand side of Eq. (3) is  
\be
\lim_{N\rightarrow\infty}\frac{1}{N}\sum_{\alpha=1}^{N}\left\{\frac{\partial \left(\frac{\widehat{\cal O} \Psi}{\Psi}\right)}{\partial c}\right\}_\alpha &=&\frac{1}{\int\Psi^{2} d\tau} \int\Psi^2\frac{\partial\left(\frac{\widehat{\cal O}\Psi}{\Psi}\right)}{\partial c} d\tau \nonumber  \\
&=&\frac{1}{\int \Psi^{2} d\tau}\int\left[\Psi \widehat{\cal O} \frac{\partial \Psi}{\partial c} - \frac{\partial \Psi}{\partial c} \widehat{\cal O} \Psi \right]d\tau \nonumber  \\
&=& 0 \nonumber
\ee

This theorem explains the simplicity of Eq. (2): the first-order change of expectation value with respect to a change of parameter comes only from the change of wave function and the Metropolis sampling weights, not from the change of the quantity (e.g. the local energy).

\section{\bf APPLICATIONS AND DISCUSSION}

To test the performance of this new analytic energy minimization scheme, a well-known trial wave function, \cite{{njr:boys},{njr:schmidt}} is used in the calculations. Explicitly, the trial wave function is expressed as 

\begin{eqnarray} 
\Psi_{\rm T}&=&D^{\uparrow}D^{\downarrow}F \nonumber \\
F &=& \exp\left(\sum_{I,i<j}U_{Iij}\right), \nonumber \\
U_{Iij} &=& \sum_{k}^{N_{I}}c_{kI}\left(\overline{r}^{m_{kI}}_{iI}\overline{r}^{n_{kI}}_{jI}+\overline{r}^{m_{kI}}_{jI}\overline{r}^{n_{kI}}_{iI}\right)\overline{r}^{o_{kI}}_{ij}, \nonumber \\
\overline{r}_{iI} &=& \frac{b_{I}r_{iI}}{1+b_{I}r_{iI}}, \nonumber \\
\overline{r}_{ij} &=& \frac{d_{I}r_{ij}}{1+d_{I}r_{ij}}, \nonumber
\end{eqnarray}
\noindent where $D^{\uparrow}$ and $D^{\downarrow}$ are the Hartree-Fock up-spin and down-spin Slater determinants in a converged STO basis set, \cite{njr:clementi} and $F$ is a positive correlation wave function. The $m_{kI}, n_{kI}$ and $o_{kI}$ are taken to be integers. All of the parameters $c_{kI}, b_{I}$ and $d_{I}$ can be optimized to obtain the lowest energy. 

With our method, a configuration size consisting of 200,000 sampling points is normally enough for satisfactory optimization for the first row atoms. Typically, one or two iterations are sufficient for convergence, requiring about fifty CPU hours on a SGI 90 MHz R8000 processor. Electrons are moved one by one with a time step chosen to maintain an acceptance ratio of 80\%. In order to generate one {\it independent} sample point, a block size of twenty sequential steps is used. 

\begin{table}[h]
\caption{Optimized ground state wave function and variational energy (with error bar and correlation energy percentage) for atoms He to C.}
\begin{tabular}{cccccccc}
$m$ & $n$ & $o$ &  He  &   Li  &   Be  &   B  &   C \\
\hline
0 & 0 & 1 &  0.2500000  &  0.2500000  &  0.2500000  &  0.2500000  &  0.2500000 \\
0 & 0 & 2 & -0.0094564  &  0.0143877  &  0.1977687  &  0.0594379  & -0.1413218 \\
0 & 0 & 3 &  0.1214671  &  0.2761786  & -0.8396261  & -0.6320118  & -0.1285105 \\
0 & 0 & 4 & -0.1399809  & -0.5225103  &  0.0634756  &  0.0444298  & -0.2202719 \\
2 & 0 & 0 &  0.2569693  & -0.0625743  & -0.3428204  & -0.2402583  & -0.1269579 \\
3 & 0 & 0 & -0.1316968  &  0.1942677  &  1.3266686  &  1.0019282  &  0.5326180 \\
4 & 0 & 0 & -0.8487197  & -0.5490759  & -2.1688741  & -1.8251190  & -1.2566210 \\
2 & 2 & 0 & -1.2608994  & -0.5235010  & -1.1187348  & -1.0333565  & -0.8918771 \\
2 & 0 & 2 &  0.8683429  &  0.6336047  &  2.1862056  &  1.9776332  &  1.6388292 \\
\hline
&Energy&(Ha)        & -2.90322(3)  & -7.47498(5)  &  -14.6413(2)  & -24.6206(3)  & -37.8054(3) \\
\hline
&Correlation&(\%)        &   99    &  93    &  72     &  73   &  75   \\
\hline
&Energy(Ref \cite{njr:schmidt})&(Ha)& -2.9029(1)  & -7.4731(6)   & -14.6332(8)& -24.6113(8)&  -37.7956(7) \\  
\hline
&Correlation(Ref \cite{njr:schmidt})&(\%)&    98    &  89    &  64    &  66   &  68   \\  
\hline
&Energy-42&(Ha)     & -2.903717(8)  & -7.47722(4)& -14.6475(1)  & -24.6257(1)& -37.8116(2)  \\
\hline
&Correlation-42&(\%)     &  100  &  98  &  79  &  77  &  79   \\
\end{tabular}
\end{table}

\begin{table}[h]
\caption{Optimized ground state wave function and variational energy (with an error bar and correlation energy percentage) for atoms N to Ne.}
\begin{tabular}{ccccccc}
      $m$ & $n$ & $o$ &  N &  O &  F &  Ne \\
\hline
0 & 0 & 1 &   0.2500000  &  0.2500000  &  0.2500000  &  0.2500000 \\
0 & 0 & 2 &  -0.2657443  & -0.3727767  & -0.4141830  & -0.4715589 \\
0 & 0 & 3 &   0.1906864  &  0.4670193  &  0.5988020  &  0.7230792 \\
0 & 0 & 4 &  -0.4252186  & -0.6653063  & -0.7861718  & -0.8802268 \\
2 & 0 & 0 &  -0.0314994  &  0.0354552  &  0.0879260  &  0.0690328 \\
3 & 0 & 0 &   0.2343842  &  0.1581261  & -0.0123869  &  0.0270636 \\
4 & 0 & 0 &  -0.9314224  & -0.8723734  & -0.6392097  & -0.6689391 \\
2 & 2 & 0 &  -0.9111045  & -1.0736302  & -1.1368462  & -1.1774526 \\
2 & 0 & 2 &   1.5219105  &  1.5985734  &  1.5418886  &  1.5606005 \\
\hline
& Energy &(Ha)&-54.5477(3)& -75.0168(1)&  -99.6792(2)& -128.8832(1)\\
\hline
& Correlation &(\%)& 78  &  80  &  84  & 86  \\
\hline
&Energy(Ref \cite{njr:schmidt})&(Ha)&-54.5390(6)&  -75.0109(4)&  -99.6685(5)& -128.8771(5) \\  
\hline
&Correlation(Ref \cite{njr:schmidt})&(\%)& 73  &  78  &  80  & 85  \\  
\hline
&Energy-42&(Ha)&-54.5563(2)&-75.0270(1)& -99.6912(2)&-128.8910(2)  \\
\hline
&Correlation-42&(\%)& 82  & 84  & 88  & 88  \\
\end{tabular}
\end{table}

To make a comparison with the variance minimization method, we choose the same set of nine parameters as Schmidt and Moskowitz \cite{njr:schmidt} with all zeroes as initial values. We also obey their constraints, enforcing the unlike-spin electron-electron cusp condition and setting $b_{I}$ and $d_{I}$ to unity. The optimized wave function and energy are shown in Tables I and II. The calculated results with our method are noticeably better for all first-row elements, especially for the so-called $2s-2p$ near-degeneracy atoms \cite{{njr:schmidt},{njr:sarsa}} Be, B and C. Approximately 10\% more correlation energy is recovered by our analytic energy derivative method. 

To demonstrate the power of our analytic energy minimization approach more fully, we optimize a forty-two parameter wave function, starting from the nine-parameter trial function discussed above. We use all terms with $m+n\le 4$ combined with $o\le 3$, $m=n=0$ with $o=4$, and all terms with $m+n>4 $ and $m\le 4$, $n\le 4$ with $o=0$. The same cusp, $b_{I}$ and $d_{I}$ constraints were obeyed. 

It is also interesting to note that in a recent VMC calculation for atoms Be, B and C, \cite{njr:sarsa} the use of additional Slater determinants enabled the authors to recover an amount of correlation energy similar to ours. Our current work demonstrates that this $2s-2p$ near-degeneracy effect for the first-row atoms accounts for less than 25\% of the correlation energy. 

\begin{table}[h]
\caption{An optimization procedure for atom C, with initial parameters as zeroes.}
\begin{tabular}{ccc}
Iteration & Energy & Error bar \\
\hline
0 & -37.68745 & 0.00039\\
1 & -37.80080 & 0.00013\\
2 & -37.80945 & 0.00012\\
3 & -37.80901 & 0.00011\\
4 & -37.80918 & 0.00011\\
\end{tabular}
\end{table}

In a typical optimization procedure with this energy derivative method, the energy value and its associated error bar decrease with the first (and possibly second) parameter moves. After that, the forces are much smaller than their error bars, indicating a local minimum. Table III shows an example of the carbon atom.

\begin{table}[h]
\caption{An optimization procedure for atom B, with optimized initial values from Ref \cite{njr:schmidt}.}
\begin{tabular}{ccc}
Iteration & Energy & Error bar \\
\hline
0 & -24.61109 & 0.00027\\
1 & -24.62044 & 0.00028\\
2 & -24.62058 & 0.00029\\
3 & -24.62043 & 0.00028\\
4 & -24.62083 & 0.00028\\
\end{tabular}
\end{table}

\begin{figure}[h]
\epsfysize=2.3 in
\centerline{\epsfbox[58 76 735 536]{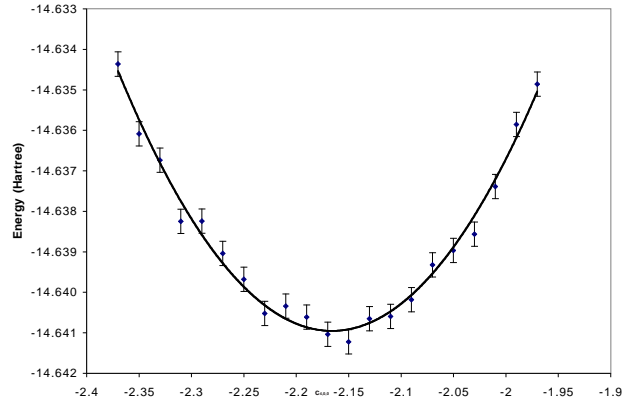}}
\caption{
Energy minimization : energies and error bars for the Be atom, as parameter for $m=4$, $n=0$, $o=0$, is varied. } 
\end{figure}

\begin{figure}[h]
\epsfysize=2.3 in
\centerline{\epsfbox[58 76 735 536]{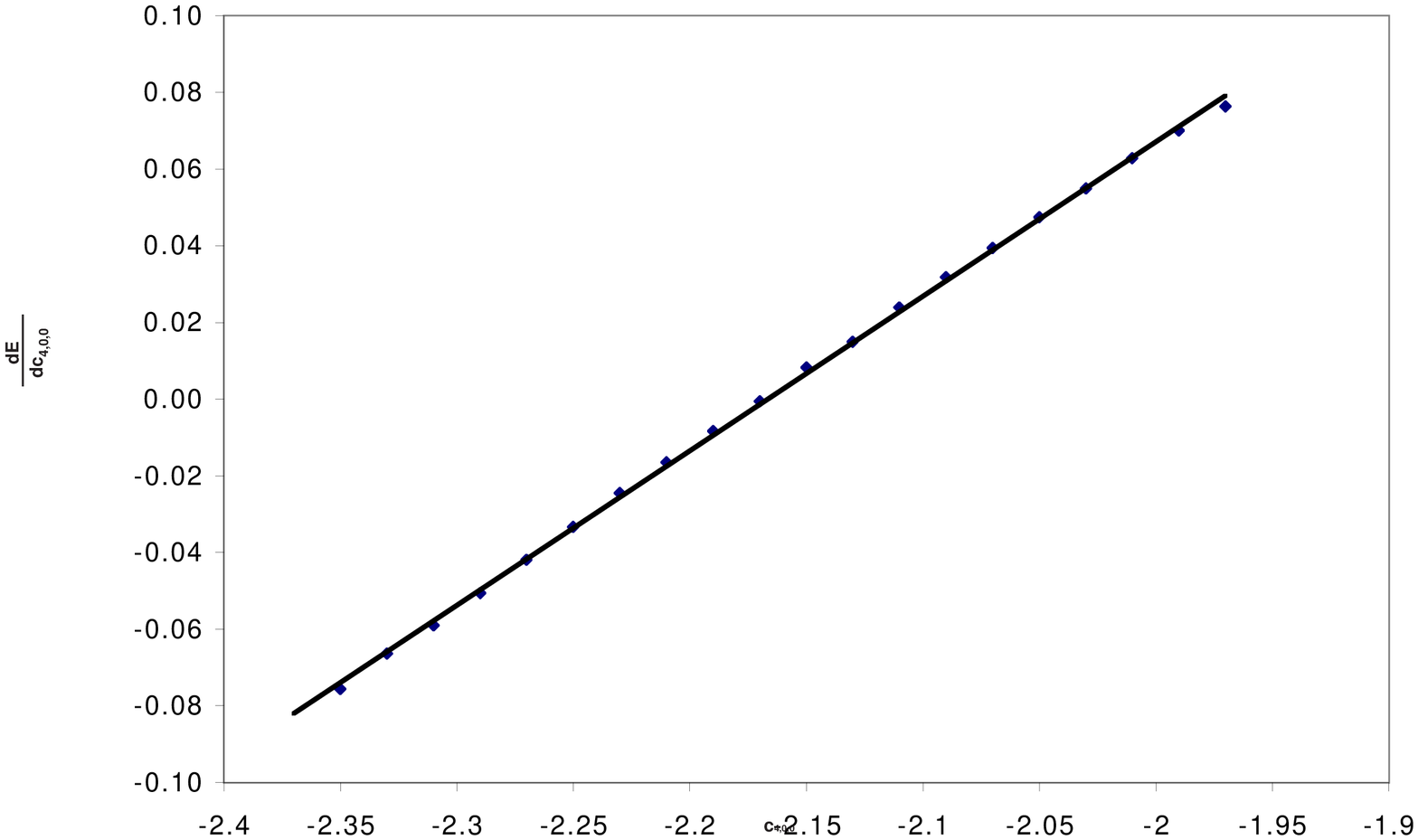}}
\caption{
Energy minimization : first-derivative of the energy with respect to the same parameter as Fig. 1.}
\end{figure}

\begin{figure}[h]
\epsfysize=2.3 in
\centerline{\epsfbox[58 76 735 536]{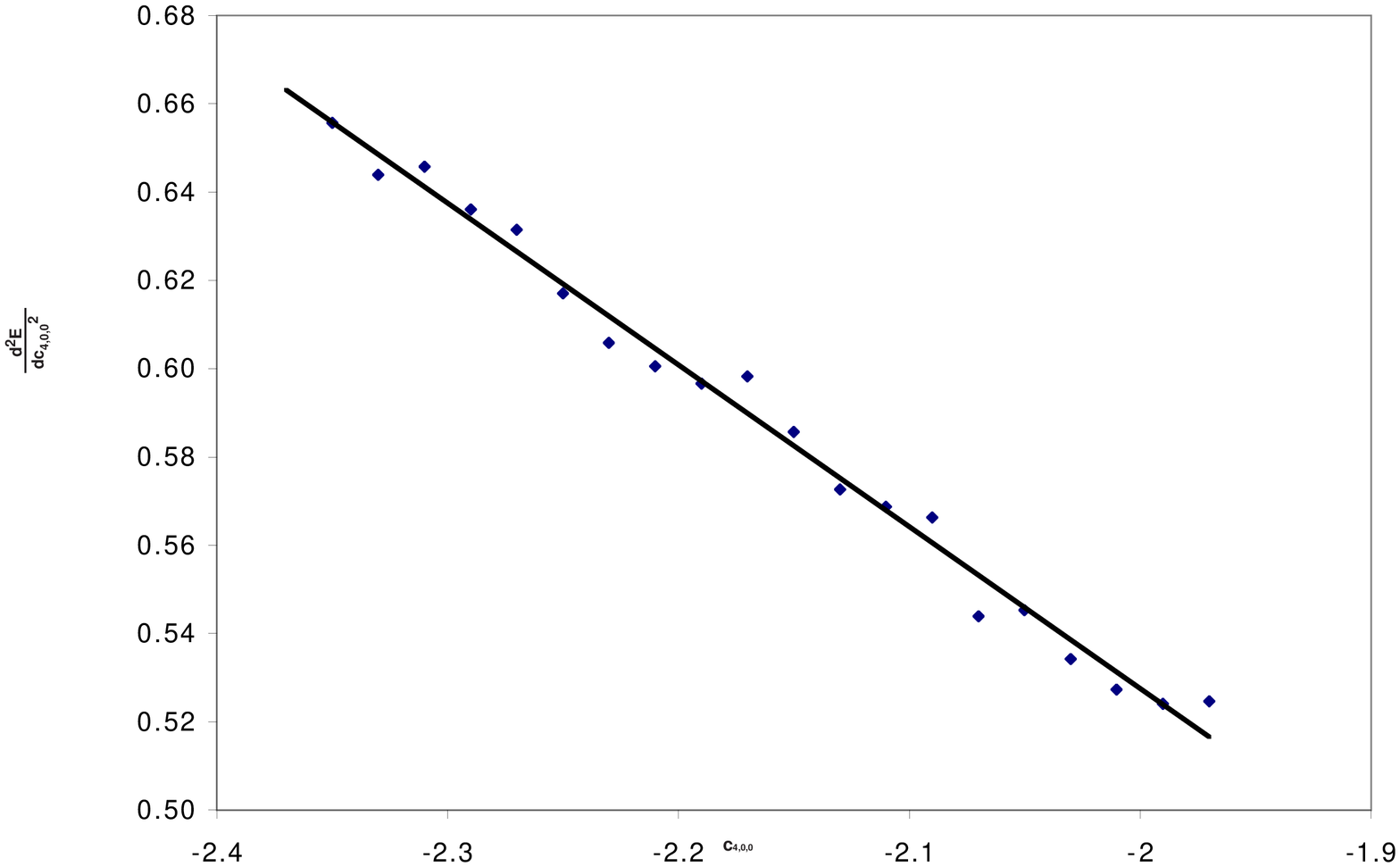}}
\caption{
Energy minimization : second-derivative of the energy with respect to the same parameter as Fig. 1. }
\end{figure}

However, rather than taking all zeroes as initial guess for variation parameters, if we start from Schmidt and Moskowitz's optimized wave function, a smaller but still sharp decrease occurs at the first iteration. Taking the atom B, for example, after one iteration, we obtained about 7\% more correlation energy. 

As one can see from Figs. 1-3, the energy derivatives are much smoother than the energy itself. As a result, it is much easier to find the parameter value which gives $dE/dc=0$ than to locate the minimum from energy data alone. As discussed in Section III, the general theorem of the local value derivatives permits reduction of noise associated with the energy derivatives for a much more efficient and reliable wave function optimization in VMC. 

After the optimization, the Hessian matrix is diagonalized to check the positivity of the eigenvalues. All of the eigenvalues are positive or small negative numbers. A positive definite Hessian guarantees all downhill movement to reach a real local minimum. The negative values are much smaller than their error bars, indicating search directions with tiny positive curvature.

\section{\bf CONCLUSIONS} 

We have explored a new method to optimize wave functions in VMC calculations. This method is a direct application of energy minimization. It is very efficient, giving quadratic convergence, and it is straightforwardly applicable to systems having a large number of parameters. In direct comparisons using identical trial wave functions, the current method yields significantly lower energy expectation values than are achieved with variance minimization for all first-row atoms.

\clearpage

\begin{table}[p]
\caption{Optimized ground state wave function and variational energy (with an error bar and correlation energy percentage) for atoms N to Ne.}
\begin{tabular}{ccccccc}
      $m$ & $n$ & $o$ &  N &  O &  F &  Ne \\
\hline
0 & 0 & 1 &   0.2500000  &  0.2500000  &  0.2500000  &  0.2500000 \\
0 & 0 & 2 &  -0.2657443  & -0.3727767  & -0.4141830  & -0.4715589 \\
0 & 0 & 3 &   0.1906864  &  0.4670193  &  0.5988020  &  0.7230792 \\
0 & 0 & 4 &  -0.4252186  & -0.6653063  & -0.7861718  & -0.8802268 \\
2 & 0 & 0 &  -0.0314994  &  0.0354552  &  0.0879260  &  0.0690328 \\
3 & 0 & 0 &   0.2343842  &  0.1581261  & -0.0123869  &  0.0270636 \\
4 & 0 & 0 &  -0.9314224  & -0.8723734  & -0.6392097  & -0.6689391 \\
2 & 2 & 0 &  -0.9111045  & -1.0736302  & -1.1368462  & -1.1774526 \\
2 & 0 & 2 &   1.5219105  &  1.5985734  &  1.5418886  &  1.5606005 \\
\hline
& Energy &(Ha)&-54.5477(3)& -75.0168(1)&  -99.6792(2)& -128.8832(1)\\
\hline
& Correlation &(\%)& 78  &  80  &  84  & 86  \\
\hline
&Energy(Ref \cite{njr:schmidt})&(Ha)&-54.5390(6)&  -75.0109(4)&  -99.6685(5)& -128.8771(5) \\  
\hline
&Correlation(Ref \cite{njr:schmidt})&(\%)& 73  &  78  &  80  & 85  \\  
\hline
&Energy-42&(Ha)&-54.5563(2)&-75.0270(1)& -99.6912(2)&-128.8910(2)  \\
\hline
&Correlation-42&(\%)& 82  & 84  & 88  & 88  \\
\end{tabular}
\end{table}

\begin{table}[p]
\caption{An optimization procedure for atom C, with initial parameters as zeroes.}
\begin{tabular}{ccc}
Iteration & Energy & Error bar \\
\hline
0 & -37.68745 & 0.00039\\
1 & -37.80080 & 0.00013\\
2 & -37.80945 & 0.00012\\
3 & -37.80901 & 0.00011\\
4 & -37.80918 & 0.00011\\
\end{tabular}
\end{table}

\begin{table}[p]
\caption{An optimization procedure for atom B, with optimized initial values from Ref \cite{njr:schmidt}.}
\begin{tabular}{ccc}
Iteration & Energy & Error bar \\
\hline
0 & -24.61109 & 0.00027\\
1 & -24.62044 & 0.00028\\
2 & -24.62058 & 0.00029\\
3 & -24.62043 & 0.00028\\
4 & -24.62083 & 0.00028\\
\end{tabular}
\end{table}

\clearpage

\end{document}